\documentclass[prl,twocolumn,showpacs]{revtex4}
\usepackage{amsfonts}
\usepackage{amsmath}
\usepackage{amssymb}
\usepackage{graphicx}

\DeclareMathOperator{\sh}{sh}
\DeclareMathOperator{\arch}{arch}
\DeclareMathOperator{\cth}{cth}
\DeclareMathOperator{\ctg}{ctg}
\DeclareMathOperator{\ch}{ch}
\input{tcilatex}

\begin{document}

\title{Diffusion in the special theory of relativity}
\author{Joachim Herrmann}
\affiliation{Max Born Institute, Max Born Stra\ss e 2a , D12489 Berlin, Germany}
\email{jherrman@mbi-berlin.de}
\pacs{05.10.Gg,
03.30.+p, 04.20.-q}

\begin{abstract}
The Markovian diffusion theory is generalized within the framework of the
special theory of relativity using a modification of the mathematical
calculus of diffusion on Riemannian manifolds (with definite metric) to
describe diffusion on Lorentzian manifolds with an indefinite metric. A
generalized Langevin equation in the fiber space of position, velocity and
orthonormal velocity frames is defined from which the generalized
relativistic Kramers equation in the phase space in external force fields is
derived. The obtained diffusion equation is invariant under Lorentz
transformations and its stationary solution is given by the J\"{u}ttner
distribution. Besides a non-stationary analytical solution is derived for
the example of force-free relativistic diffusion.
\end{abstract}

\maketitle

\section{1. Introduction}
The formulation of a consistent theory of Markovian diffusion within
the framework of the relativity theory is a long-standing problem in
physics. Over the years umpteen studies have been devoted to this
issue (see e.g. \cite{r1} - \cite{r11}) with different and
apparently irreconcilable points of view. At least up to the
knowledge of the author, a general accepted consistent solution of
this problem is still outstanding. However, besides its fundamental
theoretical interest such theory is of particular importance in
several applications such as in high energy collision experiments
(see e.g. \cite{r12}), astrophysics (see e.g. \cite{r13}) and
others.

An alternative physical approach for the description of a relativistic gas
in a heat bath is given by statistical thermodynamics and the Boltzmann
equation. J\"{u}ttner derived the thermal equilibrium distribution for a
relativistic gas already in 1911 \cite{r14}. After that many authors have
given important contributions to the development of the relativistic kinetic
theory (for an introduction see e.g. \cite{r15},\cite{r16}). In spite of
this progress in recent years a controversial debate about the correct
generalization of Maxwells velocity distribution in special relativity arose
\cite{r8}, \cite{r17}- \cite{r19}. Recently numerical microscopic
1-dimensional simulations \cite{r20} and a critical analysis of alternative
findings \cite{r21} yield arguments in favor the J\"{u}ttner distribution.
Relativistic diffusion processes has been comprehensively reviewed \cite{r11}
\ which besides the issue of stochastic relativistic diffusion theory also
includes relativistic equilibrium thermostatics and microscopic models for
Langevin-typ equations and where a more complete list of references can be
found.

Diffusion theory in the Euclidian space R$^{d}$ is a well developed
topic (see e.g. \cite{r22}, \cite{r23}). However, the description of
diffusion on a non-Euclidian manifolds M$^{d}$ is a subject
containing several pitfalls. There exists a well developed rigorous
mathematical theory of stochastic differential equations and
diffusion processes on Riemannian manifolds with a definite metric
signatur (see e.g. \cite{r24},\cite{r25}). The stochastic calculus
on Riemannian manifolds found considerable interest in mathematics
and has played a central role in recent years within the analysis in
path and loop spaces in topology and other fields. However, this
mathematical approach can not be applied to describe diffusion on
Lorentzian or Pseudo-Riemannian manifolds with an indefinite metric.
In the present paper we derive a physically motivated modification
of this calculus to describe diffusion in the phase space of the
position and velocity, in which the difficulties in the description
of diffusion on manifolds with indefinite metric signature are
bypassed. In this approach a generalized relativistic Langevin
equation in the fiber bundle of position, velocity and orthonormal
velocity frames is defined from which the generalized
non-relativistic Kramers equation in external force fields is
derived. As will be shown the derived relativistic diffusion
equation satisfies the general principle of special relativity and
is invariant under Lorentz transformations. The steady-state
solution of this equation for a heath bath with constant friction
coefficient yields the J\"{u}ttner distribution.\newline The paper
is organized as follows. In Chapter 2 the concept and main formulas
of the mathematical stochastic calculus on Riemannian manifolds is
presented. Since this approach is little-known in physics in the
appendix B some details of this calculus and its relation with the
stochastic calculus on Euclidian manifolds (appendix A) are
presented. In Chapter 3 the generalized relativistic Langevin
equation in the fiber bundle of position, velocity and orthonormal
velocity frames is defined and the relativistic diffusion equation
for the probability density function or the transition probability
is derived. In Chapter 4 the steady-state solution for a
relativistic gas in a heat bath with constant friction coefficient
and in Chapter 5 the non-steady solution for the force-free case are
derived and in chapter 6 the conclusions are presented.
\section{2. Mathematical Stochastic Calculus on Riemannian manifolds}
Stochastic differential equations in diffusion theory in a $d$-dimensional
Euclidian space R$^{d}$ with continuous pathway are defined by the
fundamental d-dimensional Wiener process W$^{a}(t).$ On a Riemannian
manifold M$^{d}$ the fundamental Wiener process is difficult to handle. By
using an inadequately posed formulation of a stochastic differntial equation
it is not assured that its solution remains on the manifold M$^{d}$ which
leads to inconsistent results. The key idea in the mathematical concept of
diffusion on general d-dimensional Riemannian manifolds M$^{d}$ (with
definite metric signature) is to define a stochastic process on the curved
manifold using the fundamental Wiener process each component of which is a
process in the Euclidian space R$^{d}$ \cite{r24},\cite{r25}. Intuitively,
we can understood this concept as follows. Consider a two dimensional
stochastic motion of a particle on a plane. If the trajectory of the
particle is traced in ink and a sphere on the plane is rolled along the
stochastic curve without slipping the resulting path which is thus
transferred defines a random curve or a stochastic Markovian process on the
sphere. This method can be applied for diffusion on a general Riemannian
manifold. The tangent space of a Riemannian manifold is endowed with an
Euclidian structure and therefore we can move the manifold in the tangent
space by construction of a parallel translation along the stochastic curve
with the help of the orthonormal frame vectors $e_{a}^{{}}=$\ $e_{a}^{i}(%
\mathbf{x})\partial _{i}$ (i,a=1...d) and the Christoffel connection
coefficients $\Gamma _{ib}^{j}$, $\mathbf{x}=(x_{1}...x_{d}),\partial
_{i}=\partial /\partial x^{i}.$ In the common mathematical denotation this
means that the random curve on the Riemannian manifold M$^{d}$ is lifted to
the horizontal curve on the frame bundle $O(M)$ and this horizontal curve
correspond uniquely to a random curve in an Euclidian space. In local
coordinates on a Riemannian manifold the infinitesimal motion of a smooth
curve c$^{i}(t)$ in M$^{d}$ is that of $\gamma ^{i}(t)$ in the tangent space
(which can be identified with R$^{d}$) by using a parallel transformation: $%
dc^{i}=e_{a}^{i}(\mathbf{x})d\gamma ^{a}$ and $de_{a}^{i}(\mathbf{x}%
)=-\Gamma _{ml}^{i}e_{a}^{l}dc^{m}$. Therefore a random curve can be defined
in the same way by using the canonical realization of a d-dimensional Wiener
process (defined in the Euclidian space) and substituting $d\gamma
^{a}\rightarrow dW^{a}(t)$. As explained in the appendix B the stochastic
differential equation describing diffusion on a Riemannian manifold in the
orthonormal frame bundle $O(M)$ with coordinates $O(M)$ =$\{x^{i},e_{a}^{i}\}
$ then is given by

\begin{eqnarray}
dx^{i}(\tau ) &=&e_{a}^{i}(\mathbf{\tau })\circ dW_{\tau }^{a}+A^{i}(\tau
)d\tau ,  \label{equ.1} \\
de_{a}^{i}(\mathbf{\tau }) &=&-\Gamma _{ml}^{i}e_{a}^{l}\circ dx^{m}(\tau )
\notag
\end{eqnarray}
Here $\eta ^{ab}e_{a}^{i}(\mathbf{x}(\tau ))e_{b}^{j}(\mathbf{x}(\tau
))=g^{ij},\partial _{i}e_{a}^{j}=-\Gamma _{ib}^{j}e_{a}^{b}$, $g^{ij}$ is
the Riemannian metric and $\eta ^{ab}=\delta ^{ab}$ the flat Euclidian
metric where $\delta ^{ab}$ is the Kronnecker symbol. The components of the
elementary Wiener process dW$^{a}$ = W$^{a}$(t+$\Delta $t)-W$^{a}$(t) are
defined in the Euclidian space with the probability density P(W$^{a})=(2D\pi
\Delta t)^{-\frac{1}{2}}\exp (-\frac{(W^{a}(t))^{2}}{2D\Delta t})$ and with
the expectation values $\langle W^{a}\rangle =0$ and\ \ \ $\langle
W^{a}(\tau )W^{b}(\tau +s)\rangle =Ds\delta _{ab}.$ A$^{i\text{ \ }}$are the
components of an arbitrary tangential vector and D is the diffusion constant
which here is independent on the time and space variables. Eq. (1) is
defined in the Stratonovich calculus (denoted by the symbol $\circ $).

Associated to each diffusion process, there is a second order differential
operator denoted as the generator $\mathbf{A}$ of the diffusion. This
operator is associated with the Kolmogorov backward equation and defined in
appendix A and B for diffusion processes on Euclidian and Riemannian
manifolds, respectively.  For the stochastic process lifted to the fiber
bundle $O(M)$ as defined in Eq.(1) the diffusion generator in the
Stratonovich integral interpretation is given by Eq.(B4) as

\begin{equation}
\mathbf{A}_{O(M)}=\frac{D}{2}\sum_{a=1}^{d}H_{a}H_{a}+H_{0}  \label{Equ.2}
\end{equation}%
where the fundamental vector fields $H_{a}$ and $H_{0}$ are given by Eq.(B3):

\begin{eqnarray}
H_{a} &=&e_{a}^{i}\frac{\partial }{\partial x^{i}}-\Gamma
_{ml}^{i}(x)e_{a}^{l}\ e_{b}^{m}\frac{\partial }{\partial e_{b}^{i}}
\label{Equ.3} \\
H_{0} &=&A^{i}(\tau ,X)\partial _{i}-\Gamma _{ml}^{i}e_{a}^{l}A^{m}(\tau )%
\frac{\partial }{\partial e_{a}^{i}}  \notag
\end{eqnarray}%
The diffusion generator $\mathbf{A}_{O(M)}$ on $O(M)$ can be projected to $%
M^{d}$ with$\ f(\mathbf{r})=f(\mathbf{x,0})$, $\mathbf{r}=(x^{i},e_{j}^{i})$
using the relation $\mathbf{A}_{O(M)}f(\mathbf{r})=\mathbf{A}_{M}f(\mathbf{x}%
)$ where

\begin{equation}
\mathbf{A}_{M}=\frac{D}{2}\sum_{a=1}^{3}(e_{a}^{i}\partial
_{i}e_{a}^{j}\partial _{j})+A^{i}\partial _{i}=(\frac{D}{2}\Delta
_{M}+A^{i}\partial _{i})  \label{Equ. 4}
\end{equation}%
and $\Delta _{M}=g^{ij}\partial _{i}\partial _{j}-g^{ij}\Gamma
_{ij}^{k}\partial _{k}$ is the Laplace-Beltrami operator on the manifold $%
M^{d}.$ The generalized Fokker-Planck equation is obtained by the adjoint of
the diffusion generator $\mathbf{A}_{M}^{\ast }$ (which includes the volume
element $\sqrt{g}$, $g=\det \{g^{ij}\}).$ Since the Laplace-Beltrami
operator is self-adjoint $\Delta _{M}=\Delta _{M}^{\ast }$ this equation
takes the form:
\begin{equation}
\frac{\partial \Phi }{\partial \tau }=-\text{div}_{x}(A\Phi )+\frac{D}{2}%
\Delta _{M}\Phi ,  \label{equ.5}
\end{equation}%
where div$_{x}(A\Phi )=g^{-\frac{1}{2}}$ $\partial _{i}$($g^{\frac{1}{2}%
}A^{i}$ $\Phi )$ is the divergence operator on the Riemannian manifold, $%
\Phi =\Phi (\mathbf{x},\tau \mid \mathbf{y},0)$ is the transition
probability with the initial condition $\Phi (\mathbf{x},0\mid \mathbf{y},0)$
$=\delta (\mathbf{x}$-$\mathbf{y})$ and adequate boundary conditions at
infinity. The probability density $\varphi $(x$^{i},\tau )$ is determined by
the same equation with the initial condition $\varphi (\mathbf{x},\tau
=0)=\varphi ^{0}(\mathbf{x})$.

A remarkable particularity of Markovian diffusion on a Riemannian
manifold is the supposition that for the diffusion coefficients in
Eq.(1) only the orthonormal frame coefficients
$e_{a}^{i}(\mathbf{x})$ are admissible which are directly related
with the geometry of the Riemannian manifold. In contrast on an
Euclidian manifold a much more general class of diffusion
coefficients is permitted.
\section{3. RELATIVISTIC DIFFUSION IN THE PHASE SPACE}
A direct application of the mathematical calculus of diffusion
processes on general Riemannian manifolds for relativistic physics
is not possible due to the supposition restricting the stochastic
formalism to the special case of a Riemannian manifold with a
definite metric signature, but in relativity theory the Lorentzian
or Pseudo-Riemannian manifolds exhibit an indefinite metric
signature (-,+,+,+). Moreover, it has been proven that Markovian
diffusion processes in the base manifold (position space) on a
Lorentzian or Pseudo-Riemannian manifold do not exist
\cite{r3},\cite{r4}. However, considering more carefully the
mathematical model described by Eq.(1)\ one can recognize a
significant difference to the physical nonrelativistic diffusion
model described by the Langevin equation. In Eq.(1) the noise term
directly acts on the position variable in the base space, while the
noise term in the Langevin equation operates like a force on the
change of the velocity in the tangent space. This crucial difference
in the mathematical model to the physically motivated Langevin
approach is the central point which enables a generalization of the
Markovian diffusion theory within the framework of the special
relativity theory performed in the phase space of coordinates
$x^{i}$ and normalized velocity variables $u^{i}$ ($i=1,2,3$). The
4-velocity space for a gas with massive particles is a hyperboloid
(or
pseudo-sphere) described by the relation (u$^{0})^{2}-$ (u$^{1})^{2}-$ (u$%
^{2})^{2}-$ (u$^{3})^{2}=1.$ This means that the relativistic velocity space
is a noncompact hyperbolic 3-dimensional Riemannian manifold (and not
Pseudo-Riemannian). Therefore with an appropriate modification we can apply
for the velocity space  the mathematical stochastic calculus for Riemannian
manifolds as presented in the appendix B. Since the stochastic force acts
directly only on the change of the velocity and not on the position
coordinates we can define the relativistic generalization of the Langevin
equation in a fiber bundle here denoted by $\emph{F}(M_{L}):$\{$x^{i}$, $%
u^{i},E_{a}^{i}\}=\emph{F}(M_{L})$ where $x^{i}$ belong to the Lorentzian
base manifold $M_{L}$, the relativistic velocity $u^{i}$ to the tangent
space $TM_{L}$ and $\ E_{a}^{i}(u)$ are the moving orthonormal frames in the
hyperbolic velocity space (this means they belong to the second order
tangent space $TTM_{L}$). Locally, this fiber bundle is simply the product
space of these three sub-spaces. With a corresponding modification of Eq.
(B1) in appendix B the generalized relativistic Langevin equations can be
defined in the fiber bundle space $\emph{F}(M_{L})$ by
\begin{eqnarray}
dx^{i}(\tau ) &=&u^{i}(\tau )d\tau ,  \notag \\
du^{i}(\tau ) &=&E_{a}^{i}(\tau ))dW^{a}(\tau )+F^{i}(\tau )d\tau ,
\label{Equ.6} \\
dE_{a}^{i}(\tau ) &=&-\gamma _{ml}^{i}(\mathbf{u})E_{a}^{l}(\tau )\circ
du^{m}(\tau )  \notag \\
&=&-\gamma _{ml}^{i}(\mathbf{u})E_{a}^{l}(\tau )[E_{b}^{m}(\tau
))dW^{b}(\tau )+F^{m}(\tau )d\tau ].  \notag
\end{eqnarray}%
Here $\tau $ is an evolution parameter along the world lines of the
particles which can be chosen as the proper time. The laboratory time $%
t=\tau u^{0}$/$c$ is a function of the proper time $\tau $ and $u^{0}$ which
here and below is defined by $u^{0}=[1+(u^{1})^{2}+(u^{2})^{2}+(u^{3})^{2}]^{%
\frac{1}{2}}$ . $\gamma _{ml}^{i}(\mathbf{u})$ are the Christoffel
connection coefficients on the hyperboloid and $F^{i}$=$K^{i}/m$, where $%
K^{i}$ are the spatial components of the 4-force, m is the rest mass of the
particles and the indices a,b denotes the spatial components in the
hyperbolic space (a,b=1,2,3). Since the stochastic force\ $dW^{a}(\tau )$ do
not act directly on the position variable $x^{i}(\tau )$ the indefinite
signature of the Lorentzian manifold here do not create any difficulty, as
it arise for a stochastic differential equation as Eq.(1) for a manifold
with indefinite metric. Sufficient conditions for the existence and
uniqueness of the stochastic differential Eq.(6) are that the drift and
diffusion coefficients satisfy the uniform Lipschitz condition and the
stochastic process $\mathbf{X}$($\tau $)=\{$\mathbf{x}$($\tau $), $\mathbf{u}
$($\tau $)\} is adapted to the Wiener process W$^{a}$($\tau )$ , that is,
the output X($\tau _{2}$) is a function of W$^{a}$($\tau _{1})$ up to that
time ($\tau _{1}\leq \tau _{2})$ \cite{r23}. The moving frames in the
hyperbolic velocity space are defined by the relation

\begin{equation}
\sum\limits_{a=1}^{3}E_{a}^{i}E_{a}^{j}=G^{ij},  \label{Equ.7}
\end{equation}%
or equivalently%
\begin{equation}
G_{ij}E_{a}^{i}E_{b}^{j}=\delta _{ab},  \label{Equ.41}
\end{equation}%
where $G_{ij}$ is the Riemannian metric of the hyperbolic velocity space, $%
G^{ij}$ the inverse matrix of $G_{ij}$ and the Christoffel connection
coefficients $\gamma _{ml}^{i}(\mathbf{u})$ on the hyperboloid are given by
\begin{equation}
\gamma _{jk}^{i}(\mathbf{u})=\frac{1}{2}G^{im}[\partial G_{jm}/\partial
u^{k}+\partial G_{mk}/\partial u^{j}-\partial G_{jk}/\partial u^{m}].
\label{Equ.8}
\end{equation}%
Since the manifold on the hyperboloid is embedded into the Minkowski space,
the metric $G_{ij}(u)$ can be calculated from the infinitesimal arc length
given by d$s_{u}^{2}=-(du^{o})^{2}+(du^{1})^{2}+(du^{2})^{2}+(du^{3})^{2}$
with $u^{0}=[1+(u^{1})^{2}+(u^{2})^{2}+(u^{3})^{2}]^{\frac{1}{2}}.$ In this
way we obtain d$s_{u}^{2}=G_{ij}(u)du^{i}du^{j}$ with $G_{ij}(\mathbf{u}%
)=\delta _{ij}-(u_{i}u_{j})/(u^{0})^{2},G=\det G_{ij}$ = ($u^{0})^{-2\text{ }%
}$ and $\gamma _{jk}^{i}(\mathbf{u})=-u^{i}G_{jk}.$ Corresponding the
definition of fundamental vector fields on $O(M)$ in the appendix B one can
now introduce the fundamental horizontal vector field $H_{a}$ and $H_{0}$ on
the fiber bundle $\emph{F}(M_{L})$. With corresponding modifications we find
from Eq.(B3):
\begin{eqnarray}
H_{a} &=&E_{a}^{i}\frac{\partial }{\partial x^{i}}-\gamma _{ml}^{i}(\mathbf{u%
})E_{a}^{l}\ E_{b}^{m}\frac{\partial }{\partial E_{b}^{i}},  \label{Equ.9} \\
H_{0} &=&u^{i}\frac{\partial }{\partial x^{i}}+F^{i}\frac{\partial }{%
\partial u^{i}}-\gamma _{ml}^{i}(\mathbf{u})E_{a}^{l}(\tau )F^{m}\frac{%
\partial }{\partial E_{a}^{i}},  \notag
\end{eqnarray}%
and the diffusion operator $\mathbf{A}_{F(M_{L})}$ \ for the stochastic
process is given as in Eq. (B4) by

\begin{equation}
\mathbf{A}_{F(M_{L})}=\frac{D}{2}\sum H_{a}H_{a}+H_{0}.  \label{Equ.10}
\end{equation}%
We project the stochastic curve from the fiber space\ $F(M_{L})$ with
coordinates $\mathbf{r}=$\{$x^{i},u^{i},E_{a}^{i}\}$ to the phase space with
coordinates \{$x^{i}$,$u^{i}\}:\mathbf{A}_{F(M_{L})}f(\mathbf{r})=\mathbf{A}%
_{P}f(\mathbf{x},\mathbf{u,0})$ , where the diffusion generator in the phase
space $\mathbf{A}_{P}$ is given by
\begin{equation}
\mathbf{A}_{P}=\frac{D}{2}\sum_{a=1}^{3}E_{a}^{i}\frac{\partial }{\partial
u^{i}}E_{a}^{j}\frac{\partial }{\partial u^{j}}+u^{i}\partial /\partial
x^{i}+F^{i}\partial /\partial u^{i}.  \label{Equ.40}
\end{equation}%
The special relativistic diffusion equation in the phase space is given by
the adjoint of the operator $\mathbf{A}_{P}$ and similar to Eq.(B7)\ the
generalized relativistic Kramers equation takes the form

\begin{equation}
\frac{\partial \Phi }{\partial \tau }=-u^{i}\frac{\partial \Phi }{\partial
x^{i}}-div_{u}(F\Phi )+\frac{D}{2}\Delta _{u}\Phi ,  \label{Equ.11}
\end{equation}%
where $\Delta _{u}$ is the Laplace Beltrami Operator of the hyperbolic
velocity space given by

\begin{eqnarray}
\Delta _{u} &=&G^{ij}\frac{\partial ^{2}}{\partial u^{i}\partial u^{j}}%
-G^{ij}\gamma _{ij}^{k}\frac{\partial }{\partial u^{k}}  \label{Equ.12} \\
&=&\frac{1}{\sqrt{G}}\frac{\partial }{\partial u^{i}}(\sqrt{G}G^{ij}\frac{%
\partial }{\partial u^{j}}),  \notag
\end{eqnarray}%
and the corresponding divergence operator is given by

\begin{equation}
div_{u}(F\Phi )=\frac{1}{\sqrt{G}}\frac{\partial }{\partial u^{i}}(\sqrt{G}%
F^{i}\Phi )  \label{Equ.13}
\end{equation}%
with $G=\det \{G_{ij}\}.$

Eq. (13) represents the relativistic generalization of the Kramers equation
for the probability density function $\Phi =\phi (\tau ;\mathbf{x},\mathbf{u}%
)$ with the initial condition $\phi (\tau =0;\mathbf{x},\mathbf{u})=\phi
_{0}(\mathbf{x},\mathbf{u})$. The transition probability is determined by
the same equation, but is defined by the initial condition $\Phi (\mathbf{x},%
\mathbf{u,0}\mid \mathbf{x}_{0},\mathbf{u}_{0},0)=\frac{1}{\sqrt{G}}\delta
(u^{1}-u_{0}^{1}$)\ \ $\delta (u^{2}-u_{0}^{2})\delta (u^{3}-u_{0}^{3}$)$%
\delta (x_{1}-x_{1}^{0}$)$\delta (x_{2}-x_{2}^{0}$)$\delta (x_{3}-x_{3}^{0}$%
). If the force F$^{i\text{ }}$depend on the time $t$ we have to substitute $%
t$ by $\ t=\tau \lbrack 1+(u^{1})^{2}+(u^{2})^{2}+(u^{3})^{2}]^{\frac{1}{2}%
}/c.$ For an external electromagnetic field $F^{\mu \nu }$ the normalized
force $F^{i}$ is given by $F^{i}=eF_{\nu }^{i}u^{\nu }$. For small
velocities ($\mid u^{i}\mid ^{2}\ll 1)$ Eq.(13)\ \ pass over to the
nonrelativistic Kramers equation \cite{r26}.

In the relativistic framework, the Lorentz-invariance of the physical laws
is one of the most fundamental property. Eq.(13) refers to a special
inertial rest frame $\Sigma $ of an observer. Now consider a second observer
at rest in an another inertial frame $\Sigma ^{`}$ that moves with constant
velocity $w$ relative to $\Sigma .$ As shown and discussed by many authors
(see e.g. \cite{r11},\cite{r16}) the probability density function $\phi
(\tau ;\mathbf{x},\mathbf{u})$ transforms as a Lorentz scalar; i.e. it
fulfills the condition

\begin{equation}
\phi `(\tau `,\mathbf{x`},\mathbf{u`})=\phi (\tau ,\mathbf{x},\mathbf{u})
\label{Equ.14}
\end{equation}%
where the variables $\tau `,\mathbf{x`},\mathbf{u`}$

\begin{equation}
x`^{i}=\Lambda _{j}^{i}x^{j}+\Lambda _{0}^{i}x^{0},\text{ }u`^{j}=\Lambda
_{j}^{i}u^{j}+\Lambda _{0}^{i}u^{0},\text{ }\tau `=\tau   \label{Equ.15}
\end{equation}%
are related by the Lorentz transformation. This requires that Eq.(13) is
invariant with respect to a Lorentz transformation. By using $x^{0}=\tau
u^{0},$ the chain rule $\partial /\partial x^{i}=\Lambda _{j}^{i}\partial
/\partial x^{`j}$ and the inverse transformation $u^{i}=\overline{\Delta }%
_{j}^{i}u`^{j}+\overline{\Delta }_{0}^{i}u`$with $\overline{\Lambda }_{\beta
}^{\alpha }\Lambda _{\gamma }^{\beta }=\delta _{\gamma \text{ }}^{\alpha }$
we find

$\ $%
\begin{eqnarray}
u^{i}\frac{\partial \phi }{\partial x^{i}} &=&u`^{i}\frac{\partial \phi }{%
\partial x`^{i}}-u^{0}\Lambda _{0}^{i}\frac{\partial \phi }{\partial x`^{i}}
\label{Equ. 16} \\
\frac{\partial \phi }{\partial \tau } &=&\frac{\partial \phi }{\partial \tau
^{`}}+u^{0}\Lambda _{0}^{i}\frac{\partial \phi }{\partial x`^{i}}  \notag
\end{eqnarray}

On a Riemannian manifold the divergence operator and the Laplace-Beltrami
operator are intrinsically invariant with respect to general coordinate
transformations. Therefore Lorentz transformation on the pseudo-sphere do
not change the explicit form of these operators. For the divergence operator
this can be simply proven by the transformation property of the covariant
differentiation $D_{j}F^{i}\equiv \partial F^{i}/\partial u_{j}+\gamma
_{jk}^{i}F^{k}$ given by $D`_{j}F`^{i}=(\partial u`^{i}/\partial
u^{k})(\partial u^{l}/\partial u`^{j})D_{l}F^{k}$ . For the divergence
operator $div_{u}(F\phi )=$ $D_{j}(F^{j}\phi )$ this yields the relation $%
D`_{j}F`^{j}$ =$D_{j}F^{j}$. The invariance of the Laplace-Beltrami operator
$\Delta _{u}\phi =$ $D_{j}$($G^{ij}\partial _{i}\phi )$\ can be similarly
proven. Since $G^{ij}\partial _{i}\phi =A^{j}$ transforms like a vector and
the divergence $D_{j}A^{j}$ is as shown above an invariant operation the
relation $\Delta _{u}\phi =\Delta _{u`}\phi $ follows. By using Eq.(18) we
express the variables $x^{i}$, $u^{j}$ \ by the new variables $x`^{i}$,$%
u`^{j}$ in the inertial frame $\Sigma ^{`}$ and account the invariance of
the divergence and Laplace-Beltrami operator. Then Eq.(13) takes the form

\begin{equation}
\frac{\partial \phi }{\partial \tau `}=-u`^{i}\frac{\partial \phi }{\partial
x`^{i}}-div_{u`}(F`\phi )+\frac{D}{2}\Delta _{u`}\phi ,  \label{Equ.17}
\end{equation}

Thus, the derived relativistic diffusion equation satisfies the general
principle of special relativity and is invariant under Lorentz
transformations. Note that Eq.(13) differs from previously derived
relativistic diffusion equations. Debbasch et al. \cite{r6} introduced a
phenomenological relativistic Langevin equation in the phase space and
derived from this a generalized Kramers equation of the classical
Ornstein-Uhlenbeck process in which the diffusion term is given by the
3-dimensional Euklidian Laplacian in the momentum space. A different
approach has been presented by Dunkel and H\"{a}nggi \cite{r8} but the final
generalized Kramers equation and its steady-state solution also differ from
Eq.(13). Both the derived equation in \cite{r6} as well those in in \cite{r8}
are not invariant under Lorentz transformations (see \cite{r11} and \cite%
{r27}).

\ For the solution of the relativistic diffusion equation it is convenient
to introduce the hyperbolic coordinate system for the 4-velocity, defined by
$u^{1}=\sh\alpha \sin \vartheta \sin \varphi ,$ $u^{2}=\sh\alpha \sin
\vartheta \cos \varphi ,$ $u^{3}=\sh\alpha \cos \vartheta $ and $u^{0}=\ch%
\alpha $. We denote the velocities in the non-Cartesian coordinates by $%
\overline{u}^{1}=\alpha $,$\overline{u}^{2}=\theta $,$\overline{u}%
^{3}=\varphi $ $,a=1,2,3.$ The metric in this coordinates are simply to
calculate and are given by $G_{11}=1,$ $G_{22}=sh^{2}\alpha
,G_{33}=sh^{2}\alpha \sin ^{2}\varphi $ and $G_{ij}=0$ for $i\neq j.$ With
the given metric the Laplace Beltrami Operator $\Delta _{u}$ in the
hyperbolic velocity space takes the form

\begin{eqnarray}
\Delta &=&\frac{\partial ^{2}}{\partial \alpha ^{2}}+2\cth\alpha \frac{%
\partial }{\partial \alpha }-  \label{Equ.18} \\
&&-\frac{1}{(\sh\alpha )^{2}}\left( \frac{\partial ^{2}}{\partial \vartheta
^{2}}+\ctg\vartheta \frac{\partial }{\partial \vartheta }+\frac{1}{(\sin
\vartheta )^{2}}\frac{\partial ^{2}}{\partial \varphi ^{2}}\right)  \notag
\end{eqnarray}%
%
%
%
%
%
%
%
%
%
%
%
%
%
%
%
%
%
%
%
%
%
%
%
%
%
%
%
%
%
%
%
%
%
%
%
%
%
%
%
%
%
%
%
%
%
and%
\begin{eqnarray}
\text{div}_{u}(F\Phi ) &=&(\sh\alpha )^{-2}\frac{\partial }{\partial \alpha }%
\left( (\sh\alpha )^{2}F^{a}\Phi \right) -  \label{Equ.19} \\
&&-(\sh\alpha )^{-1}(\sin \vartheta )^{-1}\frac{\partial }{\partial
\vartheta }\left( \sin \vartheta F^{\vartheta }\Phi \right)  \notag \\
&&-(\sh\alpha )^{-1}(\sin \vartheta )^{-1}\frac{\partial }{\partial \varphi }%
(F^{\varphi }\Phi ),  \notag
\end{eqnarray}%
is the divergence operator in the hyperbolic velocity space. Here the force
in the hyperbolic coordinate system $F^{a},F^{\vartheta },F^{\varphi }$ is
related with $F^{i}$ by $F^{a}=(\ch\alpha )^{-1}\ [\sin \vartheta (\cos
\varphi F^{1}+\sin \varphi F^{2})+\cos \vartheta F^{3}]$, $F^{\vartheta }=(%
\sh\alpha )^{-1}\ [\cos \vartheta (\cos \varphi F^{1}+\sin \varphi
F^{2})-\sin \vartheta F^{3}]\ $and $F^{\varphi }=(\sh\alpha )^{-1}\ (\sin
\vartheta )^{-1}[-\sin \varphi F^{1}+\cos \varphi F^{2}]$ with the initial
condition $\phi (\tau =0;x_{i},\alpha ,\vartheta ,\varphi )=\phi
_{0}(x_{i},\alpha ,\vartheta ,\varphi )$. The transition probability is
determined by the initial condition $\Phi (\mathbf{x},\alpha ,\vartheta
,\varphi ,0\mid \mathbf{x}_{0},\alpha _{0},\vartheta _{0},\varphi _{0},0)=(%
\sh\alpha )^{-2}(\sin \vartheta )^{-1}\delta (\alpha -\alpha _{0}$)\ \ $%
\delta (\vartheta -\vartheta _{0})\delta (\varphi -\varphi _{0}$)$\delta
(x_{1}-x_{1}^{0}$)$\delta (x_{2}-x_{2}^{0}$)$\delta (x_{3}-x_{3}^{0}$).

The diffusion of massless particles such as e.g. the diffusion of photons in
random media (see e.g.\cite{r30}) can be described in analogous way, but
with the condition ($u^{0})^{2}-(u^{1})^{2}-(u^{2})^{2}-(u^{3})^{2}=0$. For
the velocity coordinates now we choose $u^{0}=r$, $u^{1}=r\sin \vartheta
\sin \varphi ,u^{2}=r\sin \vartheta \cos \varphi ,u^{3}=r\cos \vartheta $.
The Laplace-Beltrami operator then takes the form%
\begin{equation}
\Delta =\frac{\partial ^{2}}{\partial r^{2}}+2r\frac{\partial }{\partial r}-%
\frac{1}{r^{2}}\left( \frac{\partial ^{2}}{\partial \vartheta ^{2}}+\ctg%
\vartheta \frac{\partial }{\partial \vartheta }+\frac{1}{(\sin \vartheta
)^{2}}\frac{\partial ^{2}}{\partial \varphi ^{2}}\right) .  \label{equ.20}
\end{equation}

\section{4. Steady state solution of particles in a heat bath: the J\"{u}%
ttner distribution}

First, we consider particles with a rest mass m in a gas in an isotropic
homogenous heat bath. The interaction of particles with the bath is
described by a random noise force and a friction force. In the
nonrelativistic theory the friction force is given by $f^{i}=-\nu m$v$^{i}$
where $\nu $ is the friction coefficient and v$^{i}$ are the components of
the nonrelativistic velocity. The relativistic generalization of the
friction force requires the introduction of a friction tensor $\nu _{\alpha
}^{i}$ similar to the pressure tensor in special relativity theory \cite{r6}%
, \cite{r8}. The friction force is expressed as $F^{i}=\nu _{\alpha
}^{i}[u^{\alpha }-U^{\alpha }],$ where $U^{\alpha }$ is the 4-velocity of
the heat bath. For an isotropic homogeneous heat bath the friction tensor is
given by

\begin{equation}
\nu _{\alpha }^{i}=\nu (\eta _{\beta }^{i}+u^{i}u_{\alpha }),  \label{Equ.21}
\end{equation}%
with $\nu $ denoting the scalar friction coefficient measured in the rest
frame of the particles. In the laboratory frame the heat bath is at rest
described by $U^{\alpha }=(1,0,0,0)$. Therefore the friction force is given
by $F^{i}=-\nu u^{i}u^{0}$ or in hyperbolic coordinates F$^{a}=-\nu \sh%
\alpha ,F^{\vartheta }=F^{\varphi }=0.$ We consider the spatial homogenous
and isotropic solution of equation Eq.(13) with Eq.(20) and (21) described
by
\begin{eqnarray}
\frac{\partial \phi (a)}{\partial \tau } &=&D\left[ \frac{\partial ^{2}}{%
\partial \alpha ^{2}}+2\frac{\partial }{\partial \alpha }\cth\alpha \right]
\phi (a)+  \notag \\
&&\nu (\sh\alpha )^{-2}\frac{\partial }{\partial \alpha }\left[ (\sh\alpha
)^{3}\phi (\alpha )\right] .
\end{eqnarray}%
The steady-state solution of this equation is given by $\phi (\alpha )=C\exp
\{-\chi \ch\alpha \}$ with $\chi =\nu /D$ and $C=$4$\pi K_{2}(\chi )/\chi .$
$K_{2}(\chi )$ denotes the modified Hankel function. This distribution is
identical with the J\"{u}ttner equilibrium distribution if the Einstein
relation $kT$ = $Dmc^{2}/\nu $ and consequently $\chi =$ $\frac{mc^{2}}{kT}$
is used. The above derived relativistic diffusion equation Eq.(13) yields
for the 3D case the correct thermodynamic relativistic equilibrium
distribution for a constant friction coefficient. Note that in a previously
derived relativistic diffusion equation \cite{r6} the J\"{u}ttner
equilibrium distribution for a relativist gas only arise as the steady-state
solution for a specifically adapted energy-dependent friction constant $\nu
=\nu _{0}(u^{0})^{2}.$ The generalized Kramers equation in \cite{r8} yields
for constant friction coefficients and in the Stratonovich interpretation a
modified J\"{u}ttner distribution, but using the non-standard pre-point
discretization rule the J\"{u}ttner distribution was found. Recently fully
relativistic one-dimensional molecular dynamics simulations favored the J%
\"{u}ttner distribution in the 1D case \cite{r11}.

\bigskip

\section{5. Nonsteady solution for the force-free case}

\bigskip Now we consider the unsteady solution of Eq.(13) for a spatial
homogenous gas with vanishing force F$^{i}$. We use the Laplace
transformation $\Phi (\alpha ,\vartheta ,\varphi ,\tau
)=\int\limits_{0}^{\infty }\widetilde{\Phi }(\alpha ,\vartheta ,\varphi
,\lambda )\exp (-\lambda \tau )d\lambda $ and for the eigenvalue functions
the ansatz $\ \widetilde{\Phi }_{M}^{J}(\alpha ,\vartheta ,\varphi ,\lambda
)=g_{J}^{\lambda }(\alpha )Y_{M}^{J}(\theta ,\varphi )$ where $%
Y_{M}^{J}(\theta ,\varphi )$ =P$_{M}^{J}(\vartheta )$e$^{iM\varphi }$ are
the spherical harmonics with the associated Legendre functions P$%
_{M}^{J}(\vartheta )$. For $g_{J}^{\lambda }(\alpha )$ the following
eigenvalue equation is derived:

\begin{equation}
\left\{ D\left[ \frac{\partial ^{2}}{\partial \alpha ^{2}}+2\frac{\partial }{%
\partial \alpha }\cth\alpha -\frac{J(J+1)}{\sh^{2}\alpha }\right] -\lambda
\right\} g_{J}^{\lambda }(\alpha )=0.  \label{equ.23}
\end{equation}%
Here the discrete index J takes the values $J=0,1,2,\ldots $ and $%
M=-J,-J+1,\ldots 0,1,\ldots J$. The eigenfunctions for the Laplace-Beltrami
operator with the eigenvalues $\lambda =D(\varkappa ^{2}+1)$ satisfying the
boundary condition are given by
\begin{equation}
g_{J}^{\varkappa }(z)=C_{J}^{\varkappa }\left( z^{2}-1\right) ^{J}\left(
\frac{d}{dz}\right) ^{J+1}\cos (\varkappa \arch z),  \label{equ.24}
\end{equation}%
with $z=\ch\alpha $ and $C_{J}^{\varkappa }=(-1)^{J+1}\prod\limits_{k=0}^{J}%
\frac{1}{\sqrt{2}\pi }(\varkappa ^{2}+\text{k}^{2})^{-\frac{1}{2}}$. The
eigenfunctions $\widetilde{\Phi }_{M}^{J}(\alpha ,\vartheta ,\varphi
,\varkappa )$ satisfy the relations of orthogonality and completeness. The
transition probability is determined by the initial condition $\Phi (\alpha
,\vartheta ,\varphi ,\tau =0\mid \alpha _{0},\vartheta _{0},\varphi _{0},0)=(%
\sh\alpha )^{-2}(\sin \vartheta )^{-1}\delta (\alpha -\alpha _{0}$)\ \ $%
\delta (\vartheta -\vartheta _{0})\delta (\varphi -\varphi _{0}$). Using the
orthogonality relation we can write%
\begin{eqnarray}
\Phi (\alpha ,\vartheta ,\varphi ,\tau  &\mid &\alpha _{0},\vartheta
_{0},\varphi _{0},0) \\
&=&\sum\limits_{M,J}\int\limits_{0^{{}}}^{\infty }\widetilde{\Phi }%
_{M}^{J}(\alpha ,\vartheta ,\varphi ,\varkappa )\widetilde{\Phi }_{M}^{\ast
J}(\alpha _{0},\vartheta _{0},\varphi _{0},\varkappa )  \notag \\
&&\exp [-D(\varkappa ^{2}+1)\tau ]d\varkappa .  \notag
\end{eqnarray}


Let us now consider the fundamental solution $J=0$. Substituting $%
g_{0}^{\varkappa }(\alpha )=-\frac{1}{\sqrt{2}\pi }(\sh\alpha )^{-1}\sin
\varkappa \alpha $ \ into Eq.(27)\ gives the transition probability%
\begin{multline}
\Phi (\alpha ,\tau \mid \alpha _{0},0)=C(\tau D)^{-\frac{1}{2}}\exp \{-\tau
D\}\frac{\sh\left( \frac{\alpha \alpha _{0}}{2D\tau }\right) }{\sh\alpha \sh%
\alpha _{0}}\times \\
\times \exp \left\{ -\frac{\alpha ^{2}+\alpha _{0}^{2}}{4D\tau }\right\} ,
\label{Eq.28}
\end{multline}%
with $C=2(4\pi )^{-\frac{3}{2}}$. For $\alpha _{0}\longrightarrow 0$ this
solution was first found in \cite{r28}. For small velocities ($\alpha \ll 1)$
the transition distribution \ shows a remarkable behavior. If we solve the
corresponding non-relativistic Kramers equation \cite{r26} substituting in
Eq.(25) \ $\sh\alpha \rightarrow $ $\alpha ,ch\alpha \rightarrow 1$ we find
for J=0 $g_{0}^{\varkappa }(\alpha )=-\frac{1}{\sqrt{2}\pi }(\alpha
)^{-1}\sin \varkappa \alpha $, but the eigenvalue is given by $\lambda
=D\varkappa ^{2}$ and the solution now is
\begin{multline}
\Phi (a,\tau \mid a_{0},0)=C(D\tau )^{-\frac{1}{2}}\alpha ^{-1}\alpha
_{0}^{-1}\left[ \exp \left\{ -\frac{(\alpha -\alpha _{0})^{2}}{4D\tau }%
\right\} -\right. \\
\left. -\exp \left\{ -\frac{(\alpha +\alpha _{0})^{2}}{4D\tau }\right\} %
\right] .  \label{Eq.(29)}
\end{multline}%
%
%
%
%
%
%
%
%
%
%
%
%
%
%
%
%
%
%
%
%
%
%
%
%
%
%
%
%
%
%
%
%
%
%
%
%
%
%
%
%
%
%
%
%
%
%
%
For $\alpha _{0}=0$ Eq.(29) pass to the known Wiener distribution in the
velocity space. In the limit $\alpha \ll 1$ the short time behavior of
Eq.(29) is up to an exponential small factor in agreement with Eq.(28),
however in the long time behavior both solutions differ by the exponential
factor $\exp (-D\tau ).$ In Fig.1 the relativistic distribution Eq.(28) is
presented by the solid lines for $\alpha _{0}=0$ and the Wiener distribution
Eq.(29) by the dotted lines. As can be seen both distributions differs by
orders of magnitudes even in the non relativistic region $a\ll 1$ for long
times $D\tau \gg 1$. This discrepancy can be explained by the topological
properties of the hyperbolic space\ (included by the boundary conditions)
which are different from that of the Euclidian space in the non relativistic
theory. The deep connection between local and global properties of diffusion
processes is a central topic in the mathematical field of heat kernels on
Riemannian manifolds \cite{r29}. The appearance of the factor $\exp (-D\tau
) $ can also be explained by physical arguments, it comes from that in the
hyperbolic coordinates the Jacobian is proportional to $sh^{2}\alpha $ which
is exponentially large for $\tau \rightarrow \infty .$ For large $\tau $ the
entire velocity space is explored and the small factor $\exp (-D\tau )$
cancels the exponentially large Jacobian and guarantees the probability
conservation.

Let us still compare the solution for a massive particle with that of a
massless one. With the Laplacian Eq.(22) the eigenvalues takes the form $%
\lambda =D\varkappa ^{2}$ and the eigenvalue solutions are $\widetilde{\Phi }%
_{M}^{J}(a,\vartheta ,\varphi ,\lambda )=g_{J}^{\lambda }(a)Y_{M}^{J}(\theta
,\varphi ),$ $g_{J}^{\varkappa }(z)=\sqrt{\frac{2}{\pi }}$($\frac{r}{%
\varkappa }$)$^{J}(-1)^{J}(\frac{1}{r}\frac{d}{dr})^{J}(\frac{\sin \varkappa
}{r}).$ The photon transition probability for $J=0$ then is given by%
\begin{eqnarray}
\Phi (a,\tau &\mid &a_{0},0)=(2rr_{0}\sqrt{\pi })^{-1}(\tau D)^{-\frac{1}{2}%
}sh(\frac{rr_{0}}{2D\tau })  \notag \\
&&\exp \{-\frac{r^{2}+r_{0}^{2}}{4D\tau }\}.  \label{Eq.30}
\end{eqnarray}

In comparison with Eq.28 the factor $\exp (-D\tau )$ here is absent since
the Jacobian do not increase exponentially for $\tau \rightarrow \infty .$\

\begin{figure}[tbph]
\centering
\includegraphics[width=\columnwidth]{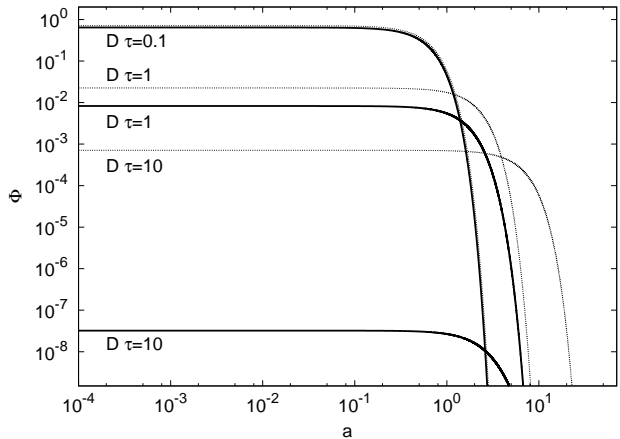}
\caption{Relativistic (solid thick lines, Eq.(28)) and non relativistic
(dotted thin lines, Eq.(29)) distributions for different times $D\protect%
\tau $ and $\protect\alpha _{0}=0$}
\label{fig:1}
\end{figure}

\section{6. Conclusions}

In conclusion, a theory of Markovian diffusion processes within the
framework of the special theory of relativity is formulated. In the
derivation of the basic relativistic diffusion equation the mathematical
calculus of stochastic differential equations on Riemannian manifolds is
used, which here is modified for the description of diffusion in the phase
space on Lorentzian manifolds with an indefinite metric. A generalized
Langevin equation in the fiber space of position, velocity and orthonormal
velocity frames is defined and the generalized relativistic Kramers equation
is derived. This diffusion equation is invariant under Lorentz
transformations. In the case of a relativistic gas in a heat bath its
steady-state solution is identical with the J\"{u}ttner distribution for
constant friction coefficients. An analytical nonsteady solution for the
transition probability is given for the special case of vanishing external
fields. This solution differs from the Wiener velocity distribution even for
small velocities due to topological reasons.

The formalism presented in this paper can be extended to a theory within the
framework of general relativity. This will be done in a forthcoming paper.

\section{Appendix A: Diffusion on Euclidian manifolds}

Using appropriate modifications the stochastic calculus on Riemannian
manifolds utilizes basic theorems and formulas from the calculus on
Euclidian manifolds. Therefore in appendix A some basic notations, formulas
and theorems for the Euclidian space are summarized including some
elementary proofs of the basic theorems.

Diffusion processes in a d-dimensional Euclidian space are described by
stochastic differential equations of the form \cite{r22}- \cite{r25}

\begin{equation}
dX^{i}=\sigma _{a}^{i}(\tau ,\mathbf{X})dW^{a}+b^{i}(\tau ,\mathbf{X})d\tau .
\tag{A1}  \label{Equ.A1}
\end{equation}%
$\mathbf{X}=(X_{1},\mathbf{...,}X_{d})$ $\in R^{d}$ is a stochastic process
with $\mathbf{X}(0)=\mathbf{x}$; $\mathbf{x=(}x_{1,...}x_{d})$ with the time
$\tau $ ($\tau \geq 0)$. The diffusion coefficients $\sigma _{a}^{i}(\tau ,%
\mathbf{X})$ are given matrices \ and the drift coefficients $b^{i}(\tau ,%
\mathbf{X})$ are coefficients of a smooth vector field. W$^{a}$ are the
components of the elementary Wiener process dW$^{a}$ = W$^{a}$(t+$\Delta $%
t)-W$^{a}$(t) with the probability density P(W$^{a})=(2\pi D\Delta \tau )^{-%
\frac{1}{2}}\exp (-\frac{(W^{a}(t))^{2}}{2D\Delta \tau })$ and the
expectation values $\langle W^{a}\rangle =0,$ \ \ \ $\langle W^{a}(\tau
)W^{b}(\tau +s)\rangle =Ds\delta _{ab},$ D is the diffusion constant.
Eq.(A1) can be transformed into an integral equation

\begin{equation}
X_{\tau }^{i}=X_{0}^{i}+\int\limits_{0}^{\tau }\sigma _{a}^{i}(s,\mathbf{X}%
)dW_{s}^{a}+\int\limits_{0}^{\tau }b^{i}(s,\mathbf{X})ds.  \tag{A2}
\label{Equ.A2}
\end{equation}

\bigskip The stochastic integral in the second term of Eq.(A2) is defined as
the limit $\int\limits_{0}^{\tau }\sigma _{a}^{i}(s,\mathbf{X}%
)dW_{s}^{a}=\sum\limits_{i=1}^{n}\sigma _{a}^{i}(s_{i}^{\ast },\mathbf{X}%
)(W^{a}(s_{i})-W^{a}(s_{i-1}))$ as $n\rightarrow \infty $. This integral
depends on the choice of the intermediate point $s_{i}^{\ast }.$ With the
choice $s_{i}^{\ast }=s_{i-1}$ (post point rule) the Ito stochastic integral
is defined. The Ito integral is a Markovian process and plays a fundamental
role in the theory of diffusion processes and most of general mathematical
treatments can only rigorously proven by using this calculus. Alternatively,
choosing $s_{i}^{\ast }=s_{i-1}$ (mid point rule) the Stratonovich
stochastic integral is defined. The Stratonovich integral has the advantage
of leading to ordinary chain rule formulas under a transformation. This
property makes the Stratonovich integral natural to use for stochastic
differential equations on Riemann manifolds. However, in general
Stratonovich integrals are not Markovian processes which hinders rigorous
mathematical treatments in most cases.

A differential equation Eq.(A1) is defined only with respect to one of the
both stochastic integrals, changing the interpretation of the integral
refers to a differential problem with different solutions. Therefore the
chosen interpretation should be denoted in the differential equation. The
symbol $\sigma _{a}^{i}(\tau ,\mathbf{X})dW^{a}$ denotes the Ito integral
interpretation and $\sigma _{a}^{i}(\tau ,\mathbf{X})\circ dW_{\tau }^{a}$
the Stratonovich interpretation.

With the Ito interpretation the solution $X_{\tau }^{i}$ of Eq.(A1) is
denoted as an Ito process if the diffusion and drift coefficients satisfy
the Lipshift condition and $\sigma _{a}^{i}(\tau ,\mathbf{X})$ is adapted to
the fundamental Wiener process $W_{\tau }^{a}$ \cite{r23}. \ An Ito process
has the important property of being Markovian. Then $\mathbf{Y}_{\tau }$ = f(%
$\mathbf{X}_{\tau }$) is also an Ito process, and its stochastic
differential equation (Ito formula) is

\begin{equation}
df=\frac{D}{2}\sum_{a=1}^{d}\{\sigma _{a}^{i}(\tau ,\mathbf{X})\sigma
_{a}^{j}(\tau ,\mathbf{X})\partial _{i}\partial _{j}f+  \notag
\end{equation}%
\begin{equation}
+b^{i}(\tau ,\mathbf{X)}\partial _{i}f\}d\tau +D\sigma _{a}^{i}(\tau ,%
\mathbf{X})\partial _{i}fdW^{a}.  \tag{A3}  \label{A3}
\end{equation}

Associated to an Ito process is the diffusion generator $\mathbf{A}$\textbf{%
\ }of $\mathbf{X}_{\tau },$which is defined to act on a suitable function $f$
by:

\begin{equation}
\mathbf{A}f=\lim_{t\rightarrow 0}\ \ \frac{E^{x}[f(\mathbf{X}_{\tau })]-f(%
\mathbf{x})}{t},  \tag{A4}  \label{Equ. A4}
\end{equation}%
where $\mathbf{x}$=$\mathbf{X}_{o}$ is the initial point of $\mathbf{X}%
_{\tau }.$ By using Eq.(A3) \ one can show that $\mathbf{A}$ is given by

\begin{equation}
\mathbf{A}f=\frac{D}{2}\sum_{a=1}^{d}\sigma _{a}^{i}(\tau ,\mathbf{X})\sigma
_{a}^{j}(\tau ,\mathbf{X})\partial _{i}\partial _{j}f+b^{i}(\tau ,\mathbf{X}%
)\partial _{i}f.  \tag{A5}  \label{A5}
\end{equation}

The generator $\mathbf{A}$ can be used in the derivation of Kolmogorov`s
backward equation. This equation describes how the expected value $E^{x}[f(%
\mathbf{X}_{\tau })]$ of any smooth function $f$ of $\mathbf{X}$ evolve in
time. If we define

\begin{equation}
u(\tau ,\mathbf{x})=E^{x}[f(\mathbf{X}_{\tau })]\ ,  \tag{A6}
\label{Equ. A6}
\end{equation}%
then $u$ satisfy the following equation:

\begin{equation}
\frac{\partial }{\partial \tau }u(\tau ,\mathbf{x})=E^{x}[\frac{d}{d\tau }f(%
\mathbf{X}_{\tau })]=\frac{d}{d\tau }E^{x}[f\mathbf{(X}_{\tau })]  \tag{A7}
\label{Equ. A8}
\end{equation}%
and with Eq.(A3)

\begin{equation}
\frac{\partial }{\partial \tau }u(\tau ,\mathbf{x})=\mathbf{A}u(\tau ,%
\mathbf{x})  \tag{A8}  \label{Equ. A9}
\end{equation}%
with $u$(0,$\mathbf{u}$)=$f$($\mathbf{x}$). The Fokker-Plack equation (or
forward Kolmogorov equation) describes how the probability density function $%
\phi (\tau ,\mathbf{x})$ of $\mathbf{X}_{\tau }$ evolve with time. This
density function is defined as a function of the d stochastic variables,
such that for any domain $\Omega $\ in the d-dimensional space of the
variables $X_{1,....,}X_{d}$, the probability that a realization of a set of
variables falls inside the domain $\Omega $ is

\begin{equation}
\Pr (X_{\tau }^{1},...X_{\tau }^{d}\in \Omega )=\int\limits_{\Omega }\phi
(\tau ,\mathbf{x})dx_{1...}dx_{d}.  \tag{A9}
\end{equation}%
The probability density function can be used to calculate the expected value
$E^{x}[f(\mathbf{X}_{\tau })]$ by $E^{x}[f(\mathbf{X}_{\tau })]$ =$%
\int\limits_{\Omega }f(\mathbf{x})\phi (\tau ,\mathbf{x})dx_{1...}dx_{d}.$
We now consider the time derivative of such expectation value and use again
Ito`s formula Eq.(A3):

\begin{equation}
E^{x}[\frac{d}{d\tau }f(\mathbf{X}_{\tau })]=\frac{d}{d\tau }E^{x}[f(\mathbf{%
X}_{\tau })]=E^{x}[\mathbf{A}f(\mathbf{X}_{\tau })]=  \notag
\end{equation}%
\begin{equation}
=\int\limits_{\Omega }\phi (\tau ,\mathbf{x})\mathbf{A}f(\mathbf{x}%
)dx_{1...}dx_{d}=  \notag
\end{equation}%
\begin{equation}
=\int\limits_{\Omega }f(\mathbf{x})\frac{\partial }{\partial \tau }\phi
(\tau ,\mathbf{x})dx_{1...}dx_{d}  \tag{A10}  \label{A10}
\end{equation}%
We integrate by parts and discard surface terms to obtain

\begin{equation}
\int\limits_{\Omega }f(\mathbf{x})\mathbf{A}^{\ast }\phi dx_{1...}dx_{d}=
\notag
\end{equation}%
\begin{equation}
=\int\limits_{\Omega }f(\mathbf{x})\frac{\partial }{\partial \tau }\phi
dx_{1...}dx_{d},  \tag{A11}
\end{equation}%
where $\mathbf{A}^{\ast }$denotes the Hermitian adjoint of $\mathbf{A}$.
Since $f$($\mathbf{x}$) is arbitrary we find for the Fokker-Plack equation
within the Ito integral interpretation:

\begin{equation}
\frac{\partial }{\partial \tau }\phi (\tau ,\mathbf{x})=A^{\ast }\phi (\tau ,%
\mathbf{x})  \tag{A12}  \label{Equ. A13}
\end{equation}%
with the adjoint operator

\begin{equation}
\mathbf{A}^{\ast }f=\frac{D}{2}\sum_{a=1}^{d}\partial _{i}\partial
_{j}\sigma _{a}^{i}(\tau ,\mathbf{X})\sigma _{a}^{j}(\tau ,\mathbf{X}%
)f-\partial _{i}b^{i}(\tau ,\mathbf{X})f.  \tag{A13}  \label{Equ. A14}
\end{equation}

Since the stochastic calculus on Riemannian manifolds is naturally
formulated in the Stratonovich integral interpretation, we will consider the
connection between both types of integrals. Let us formulate the stochastic
differential equation Eq.(A1) with the Ito interpretation by an
corresponding equation with the Stratonovich interpretation

\begin{equation}
dX^{i}=\widetilde{\sigma }_{a}^{i}(\tau ,\mathbf{X})\circ dW^{a}+\widetilde{b%
}^{i}(\tau ,\mathbf{X})d\tau .  \tag{A14}  \label{Equ. A15}
\end{equation}%
In the Stratonovich interpretation the differentiation of a function $f$ \
yields the ordinary chain rule, e.g. in the Ito formula Eq.(A3) the first
term with second-order derivatives does not appear and the above given
treatment can not be done in a consistent way. However, there exist a
connection between the Ito and the Stratonovich integrals \cite{r22}- \cite%
{r25} of functions $\varphi _{a}(\mathbf{X}(\tau ),\tau )$ in which $\mathbf{%
X}(\tau )$ is the solution of the Ito differential equation Eq.(A1):

\begin{equation}
\int\limits_{0}^{\tau }\varphi _{a}(s,\mathbf{X})dW_{s}^{a}=\int%
\limits_{0}^{\tau }\varphi _{a}(s,\mathbf{X})\circ dW_{s}^{a}-  \notag
\end{equation}%
\begin{equation}
-\frac{D}{2}\int\limits_{0}^{\tau }\sigma _{a}^{i}(\tau ,\mathbf{X})\partial
_{i}\varphi (s,\mathbf{X})\circ dW_{s}^{a}  \tag{A15}
\end{equation}

If we now make the choice

\begin{equation}
\widetilde{b}^{i}(\tau ,\mathbf{X})=b^{i}(\tau ,\mathbf{X})-\sum_{a=1}^{d}%
\frac{D}{2}\sigma _{a}^{j}(\tau ,\mathbf{X})\partial _{j}\sigma
_{a}^{i}(\tau ,\mathbf{X}),\text{ }  \notag
\end{equation}%
\begin{equation}
\sigma _{a}^{i}(\tau ,\mathbf{X}\text{)}=\widetilde{\sigma }_{a}^{i}(\tau
\mathbf{,X})  \tag{A16}
\end{equation}%
and substitute $\ \widetilde{b}^{i}(\tau ,\mathbf{X})$ into Eq.(A5), the
diffusion operator $\mathbf{A}$ in the Stratonovich interpretation is

\begin{equation}
\mathbf{A}=\frac{D}{2}\sum_{a=1}^{d}\sigma _{a}^{i}(\tau ,\mathbf{x}%
)\partial _{i}\sigma _{a}^{j}(\tau ,\mathbf{x})\partial _{j}+\widetilde{b}%
^{i}(\tau ,\mathbf{x})\partial _{i}.  \tag{A17}  \label{Equ. A18}
\end{equation}%
The Fokker-Plack equation Eq.(A12) with respect to the Stratonovich
interpretation is then given by:

\begin{equation}
\frac{\partial }{\partial \tau }\phi (\tau ,\mathbf{x})=\frac{D}{2}%
\sum_{a=1}^{d}\partial _{i}\sigma _{a}^{i}(\tau ,\mathbf{x})\partial
_{j}[\sigma _{a}^{j}(\tau ,\mathbf{x})\phi (\tau ,x)]-  \notag
\end{equation}%
\begin{equation}
-\partial _{i}\widetilde{b}^{i}(\tau ,\mathbf{x})\phi (\tau ,\mathbf{x}).
\tag{A18}
\end{equation}%
We Introduce the fundamental vector fields

\begin{equation}
L_{a}=\sigma _{a}^{i}(\tau ,\mathbf{x})\partial _{i}\text{, }L_{0}=%
\widetilde{b}^{i}(\tau ,\mathbf{x})\partial _{i}  \tag{A19}  \label{Equ. A20}
\end{equation}%
the generator\emph{\ }$\mathbf{A}$\textit{\ }of the stochastic process can
be expressed by the fundamental vector fields $L_{a}$

\begin{equation}
\mathbf{A}=\frac{D}{2}\sum_{a=1}^{d}L_{a}L_{a}+L_{0}.  \tag{A20}
\label{Equ. A21}
\end{equation}

 \section{Appendix B: Diffusion on Riemannian manifolds}

Stochastic differential equations are defined by the driving Wiener process W%
$^{a}$ ( or more general by a semi-martingale), but this process is
difficult to handle on a Riemannian manifold. In differential geometry for a
general d-dimensional Riemannian manifold M$^{d}$ (with definite metric
signature) equipped with a Christoffel connection $\Gamma _{ib}^{j}$ it is
possible to lift a smooth curve c$^{i}(t)$ in M$^{d}$ to a horizontal curve
in the tangent bundle $TM$ \ which is endowed with an Euclidian structure by
using the bundles of orthonormal frames $e_{a}^{{}}=$\ $e_{a}^{i}(x)\partial
_{i}$ (i,a=1-d). The orthonormal frame bundle $O(M$ ) is described by the
local coordinates \{$r=(x^{i},e_{j}^{i})\}=O(M)$. The infinitesimal motion
of a smooth curve x$^{i}(t)$ in M$^{d}$ is that of $\gamma ^{i}(t)$ in $O(M)$
described by the ordinary differential equations for a parallel transport%
\begin{equation}
dx^{i}=e_{a}^{i}(\mathbf{x})d\gamma ^{a},  \notag
\end{equation}%
\begin{equation}
de_{a}^{i}(\mathbf{x})=-\Gamma _{ml}^{i}e_{a}^{l}dx^{m}.  \tag{B1}
\end{equation}%
Here $\eta ^{ab}e_{a}^{i}(\mathbf{x})e_{a}^{i}(\mathbf{x})=g^{ij},\partial
_{i}e_{a}^{j}=-\Gamma _{ib}^{j}e_{a}^{b}$, $g^{ij}$ is the Riemannian metric
and $\eta ^{ab}=\delta ^{ab}$ the flat Euclidian metric where $\delta ^{ab}$
is the Kronecker symbol. $r^{i}(t)$ is called the horizontal lift of the
curve $x^{i}(t)$ to the orthonormal frame bundle $O(M$) and it lies in the
Euclidian space R$^{d+d^{2}}$. The horizontal curve $\gamma ^{i}(t)$
corresponds uniquely to a smooth curve in the tangent space (which can be
identified with an Euclidean space R$^{d}$).

Stochastic differential equations on a Riemannian manifold can be defined by
using the above described horizontal lift to the orthonormal frame bundle $%
O(M)$ endowed with an Euclidian structure \cite{r24},\cite{r25}. By using
this approach a stochastic process on the Riemannian manifold can be
constructed by using the fundamental Wiener process each component of which
is a process in the Euclidian space R$^{d}$ and interpreting the
corresponding stochastic integral in the sense of Stratonovich. This means
that the manifold is moved along a stochastic curve in the tangent space by
a parallel translation with the help of the orthonormal frame bundles and
the Christoffel connection coefficients $\Gamma _{ib}^{j}$. Correspondingly
a random curve can be defined in the same way as in Eq.(B1) by using the
canonical realization of a d-dimensional Wiener process and substituting $%
d\gamma ^{a}\rightarrow dW^{a}(t)$. Therefore the stochastic differential
equation describing diffusion on a Riemannian manifold is \cite{r24}, \cite%
{r25}

\begin{equation}
dx^{i}=e_{a}^{i}(\tau )\circ dW^{a}+A^{i}d\tau ,  \notag
\end{equation}%
\begin{equation}
de_{a}^{i}(\tau )=-\Gamma _{ml}^{i}e_{a}^{l}\ (\tau )\circ dx^{m},  \tag{B2}
\end{equation}%
where the components of an arbitrary tangential vector A$^{i\text{ \ }}$are
additionally introduced for a more general situation with account of an
external force field. The components of the elementary Wiener process dW$%
^{a} $ = W$^{a}$(t+$\Delta $t)-W$^{a}$(t) are defined in the Euclidian space
with the probability density P(W$^{a})=(2D\Delta t)^{-\frac{1}{2}}\exp (-%
\frac{(W^{a}(t))^{2}}{2D\Delta t})$ with the expectation values $\langle
W^{a}\rangle =0,$ \ \ \ $\langle W^{a}(\tau )W^{b}(\tau +s)\rangle =Ds\delta
_{ab}$.

The derivation of the Kolmogorov backward equation with the definition of
the diffusion operator $\mathbf{A}_{O(M)}$ can be performed by the same
rules as in Euclidian space in the Stratonovich calculus. Corresponding the
definition of the fundamental vector fields $L_{a}$ and $L_{0}$ \ in
Eq.(A19) one can now introduce the fundamental horizontal vector fields $%
H_{a}$ and $H_{0}$ on $O(M)$ for the extended stochastic differential system
Eq.(B2):
\begin{equation}
H_{a}=e_{a}^{i}\frac{\partial }{\partial x^{i}}-\Gamma _{ml}^{i}(\mathbf{x}%
)e_{a}^{m}\ e_{b}^{l}\frac{\partial }{\partial e_{b}^{i}},  \notag
\end{equation}%
\begin{equation}
H_{0}=A^{i}(\tau ,\mathbf{X})\partial _{i}-\Gamma _{ml}^{i}e_{a}^{l}\ (\tau
)A^{m}\frac{\partial }{\partial e_{a}^{i}},  \tag{B3}
\end{equation}%
and the operator $\mathbf{A}_{O(M)}$ for the stochastic process in the
orthonormal frame bundle is given by

\begin{equation}
\mathbf{A}_{O(M)}=\frac{D}{2}\sum_{a=1}^{d}H_{a}H_{a}+H_{0}.  \tag{B4}
\label{Equ. B4}
\end{equation}%
$\mathbf{A}_{O(M)}$ is the horizontal lift of the diffusion generator $%
\mathbf{A}_{M}$ on the manifold to the orthonormal frame bundle. Obviously,
the projection of a function in $O(M)$ to $M$ with$\ f(\mathbf{r})=f(\mathbf{%
x,0})$, $\mathbf{r}=(x^{i},e_{j}^{i})$ satisfy the relation

\begin{equation}
\mathbf{A}_{O(M)}f(\mathbf{r})=A_{M}f(\mathbf{x})  \tag{B5}  \label{Equ. B5}
\end{equation}%
where $\mathbf{A}_{M}=\frac{D}{2}\sum_{a=1}^{3}(e_{a}^{i}\partial
_{i}e_{a}^{j}\partial _{j})+A^{i}\partial _{i}=(\frac{D}{2}\Delta
+A^{i}\partial _{i})$ and $\Delta _{M}=g^{ij}\partial _{i}\partial
_{j}+g^{ij}\Gamma _{ij}^{k}\partial _{k}$ is the Laplace-Beltrami operator.
The generalized Kolmogorov backward equation on a Riemannian manifold is
obtained by

\begin{equation}
\frac{\partial }{\partial \tau }u(\tau ,\mathbf{x})=(\frac{D}{2}\Delta
_{M}+A^{i}\partial _{i})u(\tau ,\mathbf{x})  \tag{B6}  \label{Equ. B6}
\end{equation}%
As shown in the appendix A the generalized Fokker-Planck equation is given
by the adjoint of the diffusion generator $\mathbf{A}$* (which includes the
volume element $\sqrt{g}$, $g=\det \{g_{ij}\}).$ Since the Laplace-Beltrami
operator is self-adjoint $\Delta _{M}=\Delta _{M}^{\text{*}}$ \ the
generalized Fokker-Plack equation on a Riemannian manifold takes the form:
\begin{equation}
\frac{\partial \Phi }{\partial \tau }=-\text{div}_{x}(A\Phi )+\frac{D}{2}%
\Delta _{M}\Phi ,  \tag{B7}  \label{Equ. B7}
\end{equation}%
where div$_{x}(A\Phi )=g^{-\frac{1}{2}}$ $\partial _{i}$($g^{\frac{1}{2}%
}A^{i}$ $\Phi )$ is the divergence operator in the Riemannian manifold, $%
\Phi =\Phi (\mathbf{x},\tau \mid \mathbf{y},0)$ is the transition
probability with the initial condition $\Phi (\mathbf{x},0\mid \mathbf{y,}%
0)=\delta ($\textbf{x}-\textbf{y}$)$ and adequate boundary conditions at
infinity. The probability density $\varphi $($\mathbf{x},\tau )$ is
determined by the same equation with the initial condition $\varphi (\mathbf{%
x},\tau =0)=\varphi ^{0}(\mathbf{x})$.


\end{document}